\documentclass[aps,twocolumn,prd,preprintnumbers, groupedaddress, superscriptaddress, nofootinbib, 10pt]{revtex4-1}

\usepackage[hidelinks]{hyperref}
\usepackage{amssymb}
\usepackage{amsfonts}
\usepackage{graphicx}
\usepackage{epstopdf}
\usepackage{dcolumn}
\usepackage{amsmath}
\usepackage{latexsym,bm}
\usepackage{amsthm}
\usepackage{slashed}
\usepackage{float}
\usepackage{color}
\usepackage{xcolor}
\usepackage{url}
\usepackage{longtable}
\usepackage{rotating}
\usepackage{ulem}

\def \be {\begin{equation}}
\def \ee {\end{equation}}
\newcommand{\ba}{\begin{aligned}}
\newcommand{\ea}{\end{aligned}}
\def \bsp {\begin{split}}
\def \esp {\end{split}}
\def \bea {\begin{eqnarray}}
\def \eea {\end{eqnarray}}

\def\mc{\mathcal}

\def \bp{\begin{pmatrix}}
\def\ep{\end{pmatrix}}

\begin{document}

\title{5d Superconformal Field Theories and Graphs}

\author{Fabio Apruzzi}\affiliation{Mathematical Institute, University of
Oxford, Woodstock Road, Oxford, OX2 6GG, United Kingdom}
\author{Craig Lawrie}\affiliation{Department of Physics and Astronomy,
University of Pennsylvania, Philadelphia, PA 19104, USA}
\author{Ling Lin}\affiliation{Department of Physics and Astronomy, University
of Pennsylvania, Philadelphia, PA 19104, USA}
\author{Sakura Sch\"afer-Nameki}\affiliation{Mathematical Institute, University
of Oxford, Woodstock Road, Oxford, OX2 6GG, United Kingdom}
\author{Yi-Nan Wang}\affiliation{Mathematical Institute, University of
Oxford, Woodstock Road, Oxford, OX2 6GG, United Kingdom}


\begin{abstract}
\noindent We propose a graph-theoretic description to determine and
characterize 5d superconformal field theories (SCFTs) that arise as circle
reductions of  6d $\mathcal{N} = (1,0)$ SCFTs. Each 5d SCFT is captured by a
graph, called a Combined Fiber Diagram (CFD).  Transitions between CFDs encode
mass deformations that trigger flows between SCFTs. In this way, the complete
set of descendants of a given 6d theory  is obtained from a
single, marginal, CFD.  The graphs encode key physical information like the
superconformal flavor symmetry and BPS states. 
As we demonstrate for the 5d descendants of 6d minimal $(E_6, E_6)$ and $(D_k, D_k)$ conformal matter (for any $k$), our proposal not only reproduces known results, but also makes predictions in particular for thus far unknown flavor symmetry enhancements.
\end{abstract}


\maketitle

\section{Introduction}

5d $\mathcal{N}=1$ SCFTs are intrinsically non-perturbative quantum field
theories.  At low energies these can have effective descriptions in terms of
weakly coupled gauge theories, which allows one to probe certain aspects of the
SCFTs.  However, due to their strongly coupled nature, a more complete
understanding of 5d SCFTs presents a challenge that necessitates methods
beyond those of ordinary field theory, thus motivating a string-theoretic
approach. This crucially incorporates an
interpolation between the infrared (IR) and
ultraviolet (UV) fixed point. 5d theories have been engineered in string theory by
$(p,q)$-fivebrane webs \cite{Aharony:1997bh}, or M-theory on non-compact
Calabi--Yau threefolds with canonical singularities
\cite{Morrison:1996xf,Intriligator:1997pq}.  In the latter approach, there is
a particularly elegant correspondence between geometry and physics, whereby
the resolution of the singularity may be identified with an renormalization
group (RG)-flow from the UV fixed point to an effective IR description.

The case we wish to make here is that string theory does not only provide
examples, but lays out a framework to map out the full landscape of 5d SCFTs,
including a characterization of their most salient properties.  For example,
singularities in the M-theory realization, where  complex surfaces have
collapsed to points, can correspond to SCFTs.  In the smooth phase, when these
surfaces have finite volume, their geometry determines the low-energy gauge
theory descriptions for  the SCFT, if one exists.  Complex curves inside these surfaces
determine the spectrum of matter hypermultiplets charged under the gauge
algebra, as well as additional non-perturbative states.  As one approaches the
UV fixed point by collapsing the surfaces to a point, these states become part
of the BPS spectrum of the SCFT.

Recent progress in identifying M-theory geometries related to 5d SCFTs
has been made in  \cite{Jefferson:2017ahm, DelZotto:2017pti,
Jefferson:2018irk,  Apruzzi:2018nre, Bhardwaj:2018vuu, Bhardwaj:2018yhy,
Closset:2018bjz}.  The approach in this letter is fundamentally different, as
it intrinsically captures some of the strongly coupled physics and gives a
surprisingly efficient way of characterizing and mapping out the landscape of
5d SCFTs.

We define a graph, associated to each 5d SCFT,  the {\it combined fiber
diagram} (CFD), which succinctly encodes the key properties of the geometry.
Each such graph corresponds to an equivalence class of surface configurations
inside a Calabi--Yau threefold, whose singular limit defines the same SCFT.
This framework also captures UV dualities amongst distinct gauge theories.  The vertices of
each graph correspond to curves, which are contained within the surfaces, and
give rise to BPS states in the UV. 

Flows between two UV fixed points are encoded in transitions between CFDs.
These are reflected in geometric transitions that modify
the curve configuration on the surfaces, such that their collapse generates a
different singularity.  The graph theoretic description gives an efficient
method to map out all  SCFTs that can be obtained by mass deformations
starting from a given CFD (and thus SCFT).

An intrinsically strongly coupled characteristic of a 5d SCFT is its flavor
symmetry, which generally is larger than that of its low-energy  description
\cite{Seiberg:1996bd}.  Determining this enhanced flavor symmetry  is
notoriously difficult.  While techniques such as the superconformal index require an effective gauge theory
description to extract these symmetries \cite{Kim:2012gu}, these approaches are not applicable for examples without such an IR description. On the other hand, the CFD manifestly encodes
the Dynkin diagram for the superconformal flavor symmetry in terms of a marked
subgraph. The CFD-transitions correspond to precise rules how vertices are
removed and marked or unmarked.  Finally, by using the graph structure of the CFD,
we can compute the representations of BPS states under the flavor symmetry.

Our approach is rooted in the duality between M- and F-theory on a singular,
elliptically fibered Calabi--Yau threefold, $Y$.  F-theory on $Y$ determines a
6d $\mathcal{N}=(1,0)$ SCFT, with flavor symmetry $G_{6d}$, whose
$S^1$-reduction with holonomies in the 6d flavor symmetry yields 5d SCFTs
realized as M-theory on different geometric limits of $Y$.  In these limits,
we can manifestly track the unbroken subgroup of $G_{6d}$ that constitutes the
flavor symmetry \cite{Apruzzi:2018nre} and the BPS spectrum
\cite{Tian:2018icz} in 5d.  We develop the geometric foundation of this approach in
the companion paper \cite{ALLSWPartI}.  In a second companion paper
\cite{ALLSWPartII}, the focus is the gauge theory description on the
Coulomb branch of 5d SCFTs, using the methods developed in \cite{Hayashi:2014kca},
which complements the CFD approach in cases with an effective gauge theory
description.

\section{SCFTs from Graphs}\label{sec:CFD}

\noindent 
A collection of compact complex surfaces
inside a Calabi--Yau threefold defines, under
suitable assumptions
\cite{Intriligator:1997pq, Xie:2017pfl, Jefferson:2017ahm,Jefferson:2018irk}, a 5d ${\cal
N}=1$ SCFT. While a precise knowledge of the surface geometries is required to
determine an IR description, the  SCFT limit is insensitive to many of the
details of this geometry.  It is this reduced set of properties, upon which
the SCFT depends, that we encode in the  {\it Combined Fiber Diagram} (CFD):
a graph whose vertices are complex curves, $C_i$, inside the collection of 
surfaces, ${\cal S} = \bigcup_{k=1}^N S_k$, and the number of  edges connecting two
vertices $C_i$ and $C_j$ is given by the intersection number $m_{i, j}= C_i
\cdot C_j$. The integer $N$ is the rank of the 5d SCFT. 
Each vertex has labels $(n_i, g_i)$, the self-intersection number
of $C_i$ inside $\cal S$ and the genus of $C_i$ (if $g_i=0$ the label is ommitted).  A detailed derivation of 
the CFDs from the geometry is subject of the companion paper
\cite{ALLSWPartI}. 

Vertices with $(n_i,g_i) = (-2,0)$ will be marked (colored) and define a subgraph,
which  corresponds to the Dynkin diagram of the non-abelian part of the flavor group of the 5d SCFT, $G_{F}$.\footnote{We discuss here only the simply-laced case and defer the
  more general case to  \cite{ALLSWPartI}.} The rank of $G_F$ is known, as discussed anon, and
  from this one can determine the abelian factors in $G_F$. Vertices with
  $(n_i, g_i) = (-1,0)$ encode possible mass deformations.

Given a CFD a new, {\it descendant CFD}, and thereby 5d SCFT, can be constructed by a {\it (CFD-)transition}:
 remove a vertex $C_i$ with $(n_i, g_i) = (-1,0)$ and update the CFD data as follows:
 \begin{equation}\label{eq:graph_transition_rules}
  \begin{aligned}
    n_j^\prime &= n_j + m_{i,j}^2 \cr
    g_j^\prime &= g_j + \frac{m_{i,j}^2 - m_{i,j}}{2} \cr
    m_{j,k}^\prime &= m_{j,k} + m_{i,j}m_{i,k} \, ,
  \end{aligned}
\end{equation}
for $j,k \neq i$.  A marked vertex for which $n_j$ changes
becomes unmarked after the transition. 
Geometrically, a transition is the collapse of a curve $C_i$ in $\cal S$.  In the
SCFT, this corresponds to a mass deformation and subsequent RG-flow to the descendant SCFT.
Such a transition is not reversible, which reflects the nature of RG-flows, where one cannot flow ``backwards'' without knowing the correct decoupled degrees of freedom.

There are natural candidate starting points to construct descendant SCFTs, the
so-called marginal theories, whose UV fixed points are 6d $(1,0)$ SCFTs.
We define associated {\it marginal or top CFDs}, which have marked vertices forming {\it affine}
Dynkin diagrams.  Such theories and their CFDs provide the starting point,
from which our transition rules \eqref{eq:graph_transition_rules} can generate
all descendant CFDs/SCFTs.

For marginal theories, the rank of the flavor symmetry is $1 +
\text{rank}(G_{6d})$. With each transition, i.e. mass deformation, the flavor
rank drops by one, thus the superconformal flavor symmetry algebra is fully determined. 

In the present letter, we consider marginal theories originating from 6d
$(G_1,G_2)$ conformal matter (CM) theories \cite{DelZotto:2014hpa}. 
The marginal CFD contains the affine Dynkin diagram of $\widehat{G_{6d}}$ as a marked subgraph, in addition
to unmarked vertices with $(n_i, g_i) = (-1,0)$. 

\section{Rank One Theories}
\label{sec:rank_one}

\begin{figure}
\includegraphics*[height= .5 \textheight ]{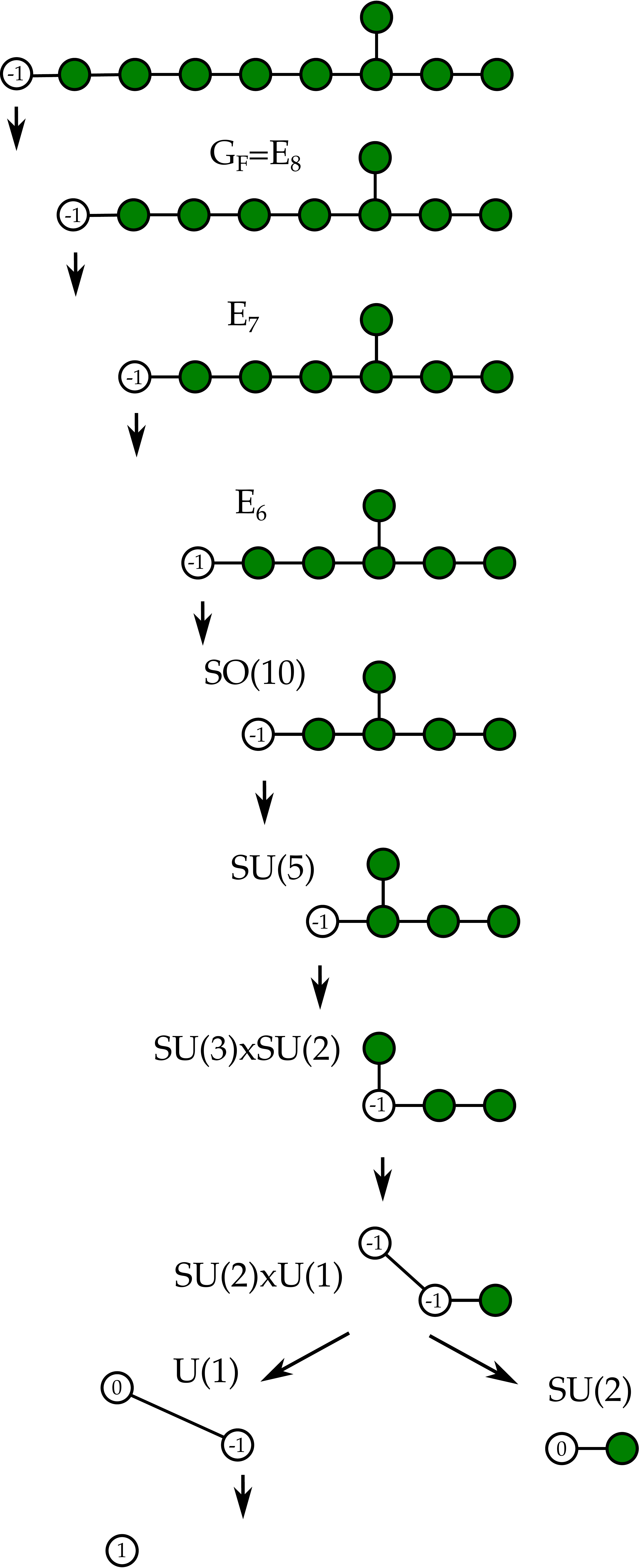}
\caption{CFD-transition tree for rank one 5d SCFTs including the superconformal flavor symmetries $G_F$.\label{fig:Rank1Tree}}
\end{figure}

In the following, we illustrate how CFD-transitions realize RG-flows between all the Seiberg and Morrison-Seiberg theories \cite{Seiberg:1996bd, Morrison:1996xf}. This results in an alternative derivation of all rank one 5d SCFTs. 
The marginal theory is associated to the rank one E-string theory and has
CFD, where the green nodes are the marked $(-2, 0)$ vertices,
\be
\includegraphics[height=0.8cm]{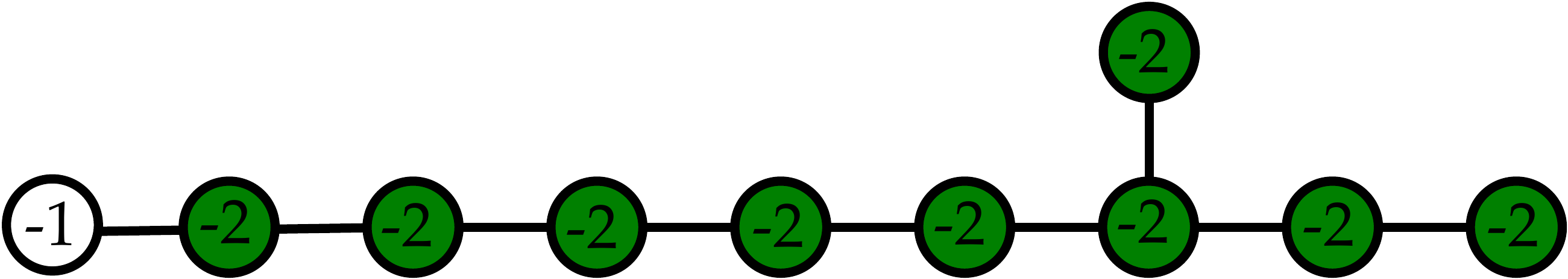} \quad .
\ee
Applying a CFD-transition to this marginal CFD describes the theory that is related by mass deformation and RG-flow. The first transition yields
\be
  \includegraphics[height= 0.8cm]{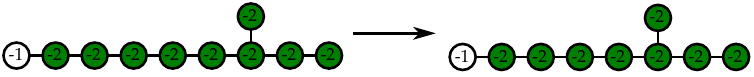} \quad ,
\ee
which is a CFD for a 5d SCFT with $E_8$ flavor symmetry. This is in fact the
UV fixed point of the $SU(2)$ theory with $N_F= 7$ fundamental flavors. The
complete tree of descendant CFDs is comprised of ten rank one 5d SCFTs with
$G_F= E_{N_F+1}$, as shown in figure \ref{fig:Rank1Tree}.  This is in agreement
with the flavor enhancement in \cite{Seiberg:1996bd, Morrison:1996xf},
including the distinction between $E_1$ and $\tilde{E}_1$, as well as
capturing the so-called ``$E_0$ theory'', which lacks a gauge description.

\section{5d SCFTs from \texorpdfstring{\boldmath{$(D_k, D_k)$}}{(Dk,Dk)} CM}\label{sec:DkDk}

Next we consider examples of arbitrary rank, descending from 6d $(D_k,D_k)$ minimal conformal matter (CM) theory on $S^1$, whose marginal CFD is
\be \label{eq:CFDD2k}
\includegraphics[height= 0.9cm]{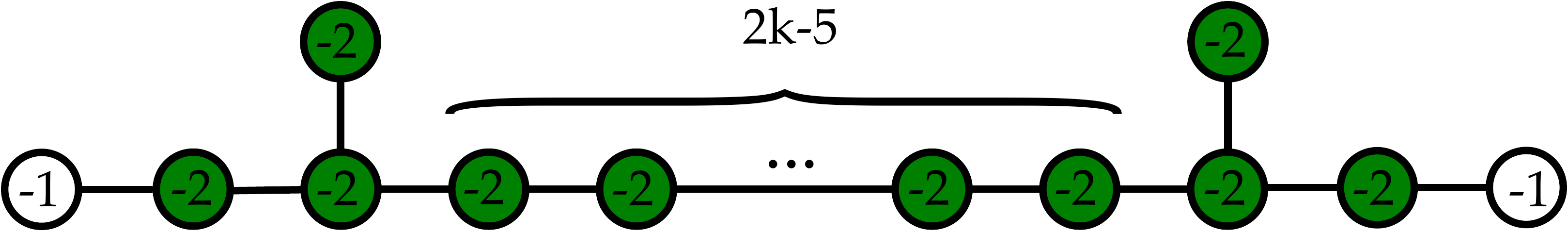} \quad .
\ee
The marked (green) $(-2)$-vertices form a $\widehat{D}_{2k}$ affine Dynkin diagram and $G_{6d}= D_{2k}$. There are $(k+2)^2-3$ descendant CFDs/SCFTs, which are shown in figure \ref{fig:DkDkAll}, including the strongly coupled flavor symmetry and spin 0 BPS states. In the supplementary material\footnote{The supplementary material is available \href{https://people.maths.ox.ac.uk/schafernamek/CFD/}{here}.} we explicitly show all descendants for $(D_9, D_9)$. 

Three dual gauge theory descriptions for the marginal theory are known
\be\label{eq:gtdescr}
\ba
SU(k-2)_0 +  2k \mathbf F \,, \quad 
 Sp(k-3)   +  &2k \mathbf F \,,\cr 
 [4 \mathbf F]  \ \hbox{--} \ SU(2)^{k-3}  \ \hbox{--} \ & [4 \mathbf F] \,,
\ea
\ee
where $SU(2)^{k-3}$ is the linear quiver with $(k-3)$ $SU(2)$ gauge nodes connected by bifundamental hypermultiplets; the factors without flavors have $\theta=0$ \cite{Bergman:2013aca, Hayashi:2015fsa}. Giving mass to the flavors, populates a subtree in figure \ref{fig:DkDkAll} of descendants that have a gauge theory description.

\begin{figure*}
  \centering
\includegraphics[height= .919\textheight]{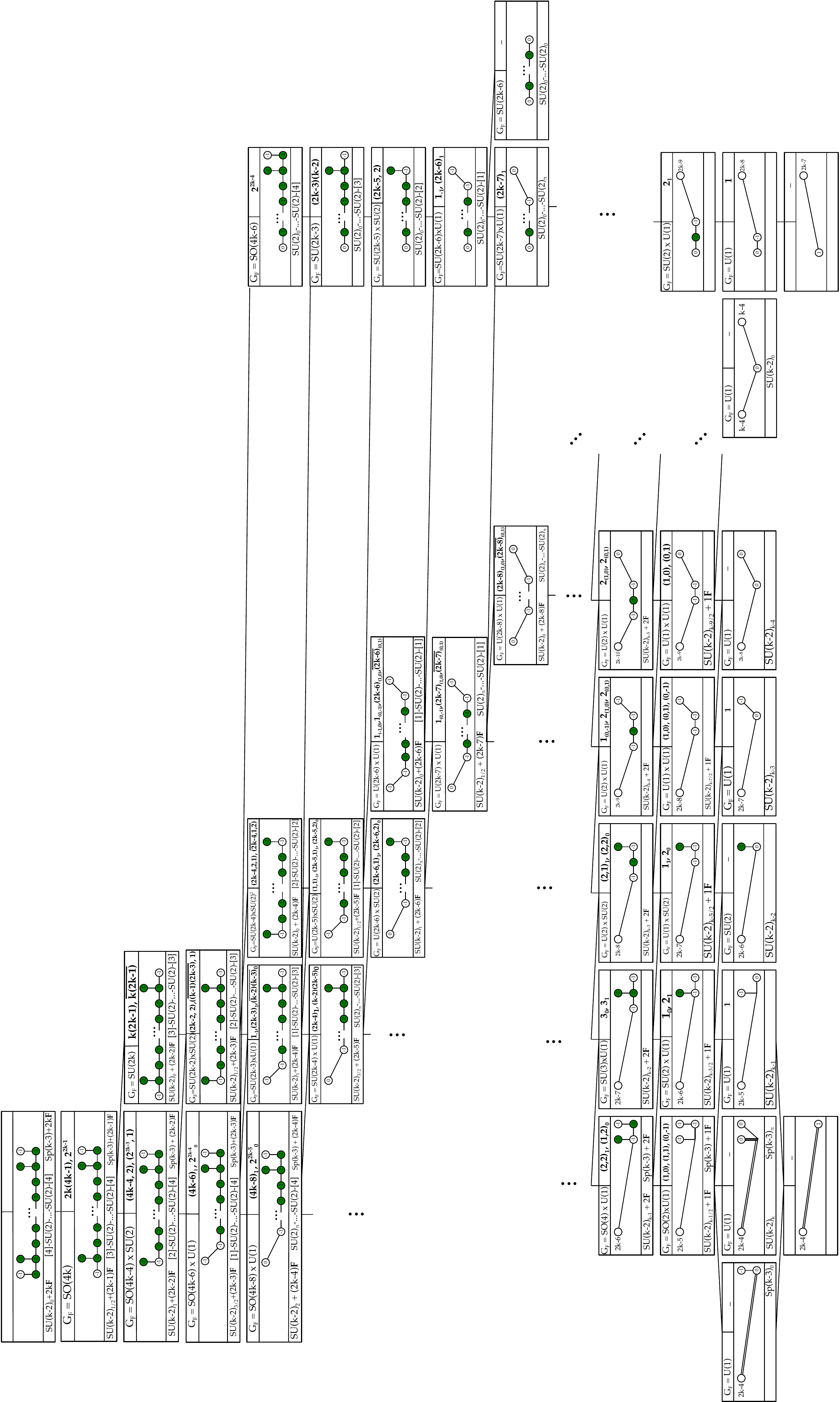}
\caption{CFDs for all 5d SCFTs descending from 6d $(D_k, D_k)$ CM. Each box contains the 5d  strongly coupled flavor symmetry, $G_F$, and the $G_F$ representations of the spin $0$ BPS states (right upper corner). In cases when there is a weakly coupled gauge theory description, this is noted at the bottom of each box. Connecting lines between boxes indicate transitions. \label{fig:DkDkAll}}
\end{figure*}

Any of the $SU(k-2)$ gauge theory descriptions is specified by the number $m$ of fundamental hypermultiplets and the Chern--Simons level, $\kappa$. Decoupling a flavor hypermultiplet shifts $\kappa$ by $\pm \frac{1}{2}$ \cite{Seiberg:1996bd}. Moreover,  $SU(k-2)_{\kappa}$ is dual to $SU(k-2)_{-\kappa}$. Overall, there are $k (k+2)$ 5d SCFTs with this weakly coupled gauge theory description. 

The CFDs predict the following superconformal flavor symmetries for theories that have an $SU(k-2)_\kappa + m  {\bf F}$ gauge theory description:
\be \label{eq:SUFS}
\ba
\kappa \quad  & \ \hbox{SCFT Flavor Symmetry $G_F$} \cr 
 k  - {m\over 2} :\ & 
\left\{\ba
				 SO(4k)  & \quad m = 2k -1\cr 
                                   SO(4k-4) \times SU(2) &\quad  m = 2k -2\cr 
                                   SO(2m) \times  U(1) & \quad  m = 0, ..., 2k-3\cr 
\ea \right.
\cr 
 k-1 - {m\over 2}:\   & 
\left\{\ba 
				SU(2k) &\quad  m = 2k -2\cr 
                                   SU(2k-2) \times  SU(2) &\quad  m = 2k -3\cr 
                                   SU(m + 1) \times  U(1) &\quad  m = 0, ..., 2k-4\cr 
\ea\right.
\cr 
 k-2 - {m\over 2} :\  & 
\left\{ \ba
SU(2k-4) \times  SU(2)^2 &\quad  m = 2k-4\cr 
                             U(m) \times  SU(2) & \quad m = 0, ..., 2k - 5\cr 
\ea\right.
\ea
\ee
These flavor symmetries agree with those  recently obtained by independent
methods in \cite{Hayashi:2015fsa,Ferlito:2017xdq,Cabrera:2018jxt}.

By decoupling stepwise the $2k$ fundamental hypermultiplets from the marginal
$Sp(k-3)$ theory in \eqref{eq:gtdescr}, we get $(2k+1)$ descendants, where the
lowest two have $Sp(k-3)_{0}$ or $Sp(k-3)_{\pi}$, and no flavors; $2k$ have a
dual $SU(k-2)$ gauge description. There is thus a unique theory with only
an $Sp(k-3)_{0}$ gauge theory description, whose classical and superconformal
flavor symmetry is $U(1)$.

For any $k$, there are six SCFTs, which have only an effective gauge theory description  in terms of the quivers
\begin{align}
\begin{split}
& SU(2)^{k-4}_0-SU(2)-[m \mathbf F],\; \;  m=1,...,4   \\
& SU(2)^{k-4}_0-SU(2)_{\theta} \,,\qquad\quad  \theta =0, \pi \,. 
\end{split}
\end{align}
The superconformal flavor symmetries are
\be
\ba
m=4  : &\quad  SO(4k-6) \cr 
m=3  :&\quad  SU(2k-3)\cr 
m=2  : & \quad SU(2k-5) \times SU(2)\cr 
m=1  : &\quad SU(2k-6) \times U(1)\cr
m=0, \; \theta=0 : & \quad SU(2k-6)\cr 
m=0, \; \theta=\pi : &\quad SU(2k-7)\times U(1) \,.
\ea
\ee
Our approach using CFDs does not only determine these flavor symmetries much
more efficiently and purely combinatorially than approaches using a gauge
theory description, we can even determine the flavor symmetry in cases when
such a description is entirely absent, i.e., the SCFT is isolated and does not
have a weakly coupled description.  In the present case, there are  $2k-6$
SCFTs that do not have any known gauge theory description, but we can determine
their superconformal flavor symmetry 
\begin{equation}
U(2k-7-i)  \,,\quad i= 0, \cdots, 2k-7  \,.
\end{equation}
These  CFDs and their associated geometries \cite{ALLSWPartI} are evidence
that such non-trivial 5d UV fixed points exist;  these have been observed for
rank two, i.e. $k=5$, in \cite{ALLSWPartI, Jefferson:2018irk, Hayashi:2018lyv}.

\section{5d SCFTs from \texorpdfstring{\boldmath{$(E_6,E_6)$}}{(E6,E6)} CM}

Another class of higher rank theories, that have thus far not been studied in
generality, are the SCFTs descending from the $(E_6,E_6)$ minimal CM theory, which are
rank five.  The marginal CFD is 
\be \label{eq:CFDE6}
\includegraphics[height= 1.5cm]{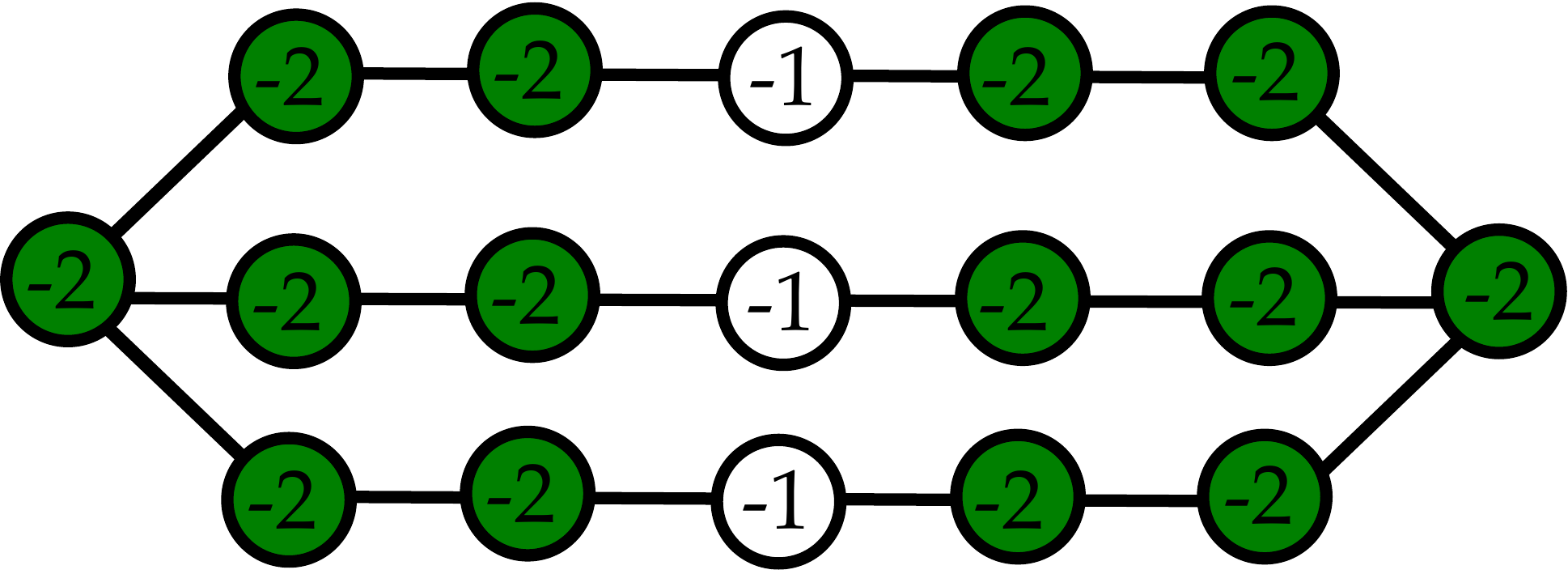} \quad .
\ee
CFD-transitions applied to this yield 207 descendant CFDs/SCFTs, which are attached in the supplementary Mathematica notebook. This predicts a large class of new 5d SCFTs. The only known weakly coupled description of the marginal theory is the quiver \cite{DelZotto:2014hpa}
\begin{equation} \label{eq:E6E6Q}
 [2]-SU(2)-\overset{%
\begin{array}
[c]{c}%
[2]\\
|\\
SU(2)\\
|
\end{array}
}{SU(3)_0}-SU(2)-[2] \,.
\end{equation}

Decoupling the flavor hypermultiplets of each $SU(2)$, step-by-step, yields
descendants with quiver descriptions. As a shorthand, we denote these by a triple
$(q_1, q_2, q_3)$, where the $q_i$ is either the number of fundamentals under,
or the theta angle of, each of the three $SU(2)$ factors in the quiver. For these quivers we find 
the following superconformal flavor symmetries: 
\begin{equation}
  \begin{aligned}
    (1{\bf F}, 2{\bf F}, 2{\bf F}) \, &: \, E_6 \times E_6 \cr
    (0, 2{\bf F}, 2{\bf F}), \ (\pi, 2{\bf F}, 2{\bf F}) \, &: \, E_6 \times SU(6) \cr
    (1{\bf F}, 1{\bf F}, 2{\bf F}) \, &: \, SO(10)^2 \times U(1) \cr
    (0, 1{\bf F}, 2{\bf F}),\  (\pi, 1{\bf F}, 2{\bf F}) \, &: \, SO(10) \times SU(5) \times U(1) \cr
    (1{\bf F}, 1{\bf F}, 1{\bf F}) \, &: \, SO(8)^2 \times U(1)^2 \cr
    (0, 0, 2{\bf F}), \  (\pi, \pi, 2{\bf F}) \, &: \, SO(10) \times SU(4) \times U(1) \cr
    (0, \pi, 2{\bf F}) \, &: \, SU(5)^2 \times U(1) \cr
    (0, 1{\bf F}, 1{\bf F}) , \ (\pi, 1{\bf F}, 1{\bf F}) \, &: \, SO(8) \times SU(4) \times U(1)^2 \cr
    (0, 0, 1{\bf F}), \ (\pi, \pi, 1{\bf F}) \, &: \, SO(8) \times SU(3) \times U(1)^2 \cr
    (0, \pi, 1{\bf F}) \, &: \, SU(4)^2 \times U(1)^2 \cr
    (0, 0, 0) ,  (\pi, \pi, \pi) \, &: \, SO(8) \times SU(2) \times U(1)^2 \cr
    (0, 0, \pi), \ (\pi, \pi, 0) \, &: \, SU(4) \times SU(3) \times U(1)^2 \,.
  \end{aligned}
\end{equation}
This populates only a small subtree of 12 elements in the CFD tree.  It is notable that the
CFDs are sensitive to the number of independent discrete parameters, e.g.,
they capture dualities between theories with different
theta angles \cite{Zafrir:2014ywa, Yonekura:2015ksa}. It would be
interesting to determine the gauge theory
descriptions, where they exist, for the remaining 195
CFD/SCFTs.

\section{BPS States}\label{sec:BPS}

BPS states, $\Phi_C$, of 5d gauge theories arise in M-theory from wrapped
M2-branes on holomorphic curves $C$ in ${\cal S}$. We
consider curves $C$ with genus $g(C)=0$ here, where $\Phi_C$
 transform under the 5d massive little group $SO(4)$ as  \cite{Witten:1996qb, Gopakumar:1998jq}
\begin{equation}
  R_n=\left(\frac{n}{2},\frac{1}{2}\right)\oplus 2\left(\frac{n}{2},0\right) \,,
\end{equation}
where $n$  is the dimension of the moduli space $\mc{M}_C$ of $C$. 
We will compute states with $n=0$ and refer to them as `spin 0'. 

In the language of CFDs, the curve $C$ is a non-negative linear combination of the curves (i.e., vertices)
shown in the CFD. 
The genus and self-intersection number is determined by recursively applying
\begin{equation}
  \begin{aligned}
    (C_1 + C_2)^2 &= C_1^2 + C_2^2 + 2 C_1 \cdot C_2 \cr
    g(C_1 + C_2) &= g(C_1) + g(C_2) + C_1 \cdot C_2 - 1 \,.
  \end{aligned}
\end{equation}
We enumerate  the $g=0$, spin 0 BPS states in terms of curves with $C \cdot C = -1$. We write
\begin{equation}
  C = \sum_i q_i C_i \,,
\end{equation}
where $C_i$ are the nodes in the CFD, and $q_i \geq 0$, which are constrained by the genus and self-intersection
number of $C$. Each curve is associated to a weight of a
representation of the flavor symmetry, where the highest weights 
under the non-abelian subalgebra, $H_F$, are determined through the intersection numbers between $C$ and the
marked curves, $F_i$,  in the CFD, by requiring
\be
C\cdot F_i\geq 0\ ,\ (i=1,\dots,\mathrm{rk}(H_F))\,.
\ee
The charges under the abelian subalgebra are
determined through the intersection with specific
combinations of unmarked vertices orthogonal to $H_F$, the $U(1)$ generators.

Applying this to the Seiberg theories reproduces the spin 0
BPS states in \cite{Huang:2013yta}.
For the $(D_k, D_k)$ descendants, the spin $0$ states are listed in figure
\ref{fig:DkDkAll}, which are predictions for BPS states in 5d strongly coupled
SCFTs.

\section*{Acknowledgements}

We thank J.~Distler, J.~J.~Heckman, N.~Mekareeya, A.~Tomasiello, 
G.~Zafrir and in particular M.~Weidner for discussions.  The work of FA, SSN, YNW is supported by the ERC
Consolidator Grant 682608 ``Higgs bundles: Supersymmetric Gauge Theories and
Geometry (HIGGSBNDL)''.  CL is supported by NSF CAREER grant PHY-1756996; LL
is supported by DOE Award DE-SC0013528Y.  FA and CL thank the 2019 Pollica
summer workshop, where part of this work was completed. YNW thanks the Aspen
Center for Physics, which is supported by National Science Foundation grant
PHY-1607611, where part of this work was finished.  SSN thanks the Mainz
Institute for Theoretical Physics, Milano Bicocca University and Kavli IPMU
hospitality during the completion of this work.  YNW was also partially
supported by a grant from the Simons Foundation at the Aspen Center for
Physics.  The authors thank the 2019 String-Phenomenology Conference and CERN
for hospitality during the final stages of this work.


%

\end{document}